\documentstyle[12pt]{article}
\def\ie{\hbox{\it i.e.}{}}      
      \def\cf{\hbox{\it cf.}{}}

\catcode`\@=11

\@addtoreset{equation}{section}

\def\@normalsize{\@setsize\normalsize{15pt}\xiipt\@xiipt
\abovedisplayskip 14pt plus3pt minus3pt%
\belowdisplayskip \abovedisplayskip
\abovedisplayshortskip  \z@ plus3pt%
\belowdisplayshortskip  7pt plus3.5pt minus0pt}
\def\small{\@setsize\small{13.6pt}\xipt\@xipt
\abovedisplayskip 13pt plus3pt minus3pt%
\belowdisplayskip \abovedisplayskip
\abovedisplayshortskip  \z@ plus3pt%
\belowdisplayshortskip  7pt plus3.5pt minus0pt
\def\@listi{\parsep 4.5pt plus 2pt minus 1pt
            \itemsep \parsep
            \topsep 9pt plus 3pt minus 3pt}}

\def\underline#1{\relax\ifmmode\@@underline#1\else
        $\@@underline{\hbox{#1}}$\relax\fi}
\@twosidetrue
\relax

\catcode`@=12

\catcode`\@=11

\def\section{\@startsection{section}{1}{\z@}{3.5ex plus 1ex minus
   .2ex}{2.3ex plus .2ex}{\large\bf}}

\def\ps@headings{\def\@oddfoot{}\def\@evenfoot{}
\def\@oddhead{\hbox{}\hfill
        \makebox[.5\textwidth]{\raggedright\ignorespaces --\thepage{}--
        \hfill }}
\def\@evenhead{\@oddhead}
\def\subsectionmark##1{\markboth{##1}{}}
}

\ps@headings

\catcode`\@=12

\relax

\def\figcap{\section*{Figure Captions\markboth
        {FIGURECAPTIONS}{FIGURECAPTIONS}}\list
        {Fig. \arabic{enumi}:\hfill}{\settowidth\labelwidth{Fig. 999:}
        \leftmargin\labelwidth
        \advance\leftmargin\labelsep\usecounter{enumi}}}
 \relax
\def\tablecap{\section*{Table Captions\markboth
        {TABLECAPTIONS}{TABLECAPTIONS}}\list
        {Table \arabic{enumi}:\hfill}{\settowidth\labelwidth{Table 999:}
        \leftmargin\labelwidth
        \advance\leftmargin\labelsep\usecounter{enumi}}}
 \relax
\def\reflist{\section*{References\markboth
        {REFLIST}{REFLIST}}\list
        {[\arabic{enumi}]\hfill}{\settowidth\labelwidth{[999]}
        \leftmargin\labelwidth
        \advance\leftmargin\labelsep\usecounter{enumi}}}
 \relax

\catcode`\@=11

\def\ps@headings{\def\@oddfoot{}\def\@evenfoot{}
\def\@oddhead{\hbox{}\hfill
        \makebox[.5\textwidth]{\raggedright\ignorespaces --\thepage{}--
        \hfill }}
\def\@evenhead{\@oddhead}
\def\subsectionmark##1{\markboth{##1}{}}
}

\ps@headings

\relax

\def\firstpage#1#2#3#4#5#6{
\begin{document}
\begin{titlepage}
\nopagebreak
\title{\begin{flushright}
        \vspace*{-1.8in}
        {\normalsize CERN--TH/95-300 -- hep-th/9511108}\\[-9mm]
   {\normalsize CPTH--S396.1195 -- LPTENS--95/49}\\[7mm]
\end{flushright}
\vfill
{#3}}
\author{\large #4 \\[.2cm] #5}
\maketitle
\vskip -7mm
\nopagebreak
\begin{abstract}
{\noindent #6}
\end{abstract}
\vfill
\begin{flushleft}
\rule{16.1cm}{0.2mm}\\[-3mm]
$^{\star}${\small Research supported in part by\vspace{-4mm}
the National Science Foundation under grant
PHY--93--06906, \linebreak in part by the EEC contracts \vspace{-4mm}
SC1--CT92--0792,
CHRX-CT93-0340 and SC1*CT92--0789, \linebreak in part by the U.S. DOE grant
DE-FG03-91ER40662 Task C, \vspace{-4mm} and in part by
CNRS--NSF \linebreak grant INT--92--16146.}\\[-3mm]
$^{\dagger}${\small Laboratoire Propre du CNRS UPR A.0014.}\\[-3mm]
$^{\natural}${\small Unit\'e Propre du CNRS, associ\'ee \`a l'Ecole Normale
Sup\'erieure et \`a l'Universit\'e de Paris-Sud.}
November 1995
\end{flushleft}
\thispagestyle{empty}
\end{titlepage}}
\newcommand{\dal}{\raisebox{0.085cm}
{\fbox{\rule{0cm}{0.07cm}\,}}}
\newcommand{\dt}{\partial_{\langle T\rangle}}
\newcommand{\dtbar}{\partial_{\langle\bar{T}\rangle}}
\newcommand{\al}{\alpha^{\prime}}
\newcommand{\mst}{M_{\scriptscriptstyle \!S}}
\newcommand{\mpl}{M_{\scriptscriptstyle \!P}}
\newcommand{\dv}{\int{\rm d}^4x\sqrt{g}}
\newcommand{\lv}{\left\langle}
\newcommand{\rv}{\right\rangle}
\newcommand{\ph}{\varphi}
\newcommand{\sbar}{\,\bar{\! S}}
\newcommand{\xbar}{\,\bar{\! X}}
\newcommand{\fbar}{\,\bar{\! F}}
\newcommand{\zbar}{\bar{z}}
\newcommand{\dbar}{\,\bar{\!\partial}}
\newcommand{\tbar}{\bar{T}}
\newcommand{\taubar}{\bar{\tau}}
\newcommand{\ubar}{\bar{U}}
\newcommand{\ybar}{\bar{Y}}
\newcommand{\phb}{\bar{\varphi}}
\newcommand{\cm}{Commun.\ Math.\ Phys.~}
\newcommand{\pr}{Phys.\ Rev.\ D~}
\newcommand{\pl}{Phys.\ Lett.\ B~}
\newcommand{\ibar}{\bar{\imath}}
\newcommand{\jbar}{\bar{\jmath}}
\newcommand{\np}{Nucl.\ Phys.\ B~}
\newcommand{\e}{{\rm e}}
\newcommand{\be}{\begin{equation}}
\newcommand{\en}{\end{equation}}
\newcommand{\gsi}{\,\raisebox{-0.13cm}{$\stackrel{\textstyle
>}{\textstyle\sim}$}\,}
\newcommand{\lsi}{\,\raisebox{-0.13cm}{$\stackrel{\textstyle
<}{\textstyle\sim}$}\,}
\date{}
\firstpage{3118}{IC/95/34}
{\large\bf $N=2$ Heterotic Superstring and its Dual Theory\\[-4mm]
in Five Dimensions$^{\star}$} {I. Antoniadis$^{\,a}$, S. Ferrara$^{b}$
and T.R. Taylor$^{\,c,d}$}
{\normalsize\sl
$^a$Centre de Physique Th\'eorique, Ecole Polytechnique,$^\dagger$
{}F-91128 Palaiseau, France\\[-3mm]
\normalsize\sl
$^b$Theory Division, CERN, 1211 Geneva 23, Switzerland\\[-3mm]
\normalsize\sl
$^c$Laboratoire de Physique Th\'eorique de l'Ecole Normale
Sup\'erieure,$^\natural$\\[-5mm]\normalsize\sl
24, rue Lhomond, F-75231 Paris, France\\[-3mm]
\normalsize\sl $^d$Department of Physics, Northeastern
University, Boston, MA 02115, U.S.A.}
{We study quantum effects in five dimensions in heterotic superstring
theory compactified on $K_3\times S_1$ and analyze the conjecture that
its dual effective theory is eleven-dimensional supergravity
compactified on a Calabi-Yau threefold. This theory is also
equivalent to type II superstring theory compactified on the same
Calabi-Yau manifold, in an appropriate large volume limit. In
this limit the conifold singularity disappears and is replaced by a
singularity associated to enhanced gauge symmetries, as na\"{\i}vely
expected from the heterotic description. Furthermore, we exhibit the existence
of additional massless states which appear in the strong coupling regime
of the heterotic theory and are related to a different type of
singular points on Calabi-Yau threefolds.}
\section{Introduction}

Among different duality conjectures, one of the most appealing is the
possible relation of superstring theory with an underlying 11-D theory
which may have only a non-perturbative definition \cite{w}. There
is an evidence that this theory, whose low-energy limit is described by
the coupling constant-free 11-D supergravity, must be suitably
implemented by two- and five-brane solitonic solutions, in order to be
at least  equivalent to lower dimensional string theories
\cite{branes}. These solutions imply that in seven dimensions the
heterotic string  compactified on $ T_3$ should be equivalent to 11-D
supergravity compactified on $K_3$ \cite{w} and that in five dimensions the
heterotic string compactified on $K_3\times S_1$ should be equivalent
to 11-D supergravity compactified on a Calabi-Yau threefold \cite{cf,pt}.
These compactifications preserve $N=2$ simple supersymmetry in $D=7$ and
in $D=5$, respectively.\footnote{We call simple the smallest possible
supersymmetry existing in a given dimension $D$.}

The interest in dealing with a five-dimensional theory is that, unlike in
higher
dimensions, the spectrum of massless states can change model by model.
The numbers of neutral massless vector multiplets and hypermultiplets
of the heterotic string, in the abelian phase, should be in
correspondence with the Hodge numbers $h_{(1,1)}$ and $h_{(2,1)}$ of
the 11-D theory \cite{d,ht}. Indeed, in $D=5$, the number of vector
multiplets is $n_V=h_{(1,1)}-1$ and the number of hypermultiplets
$n_H=h_{(2,1)}+1$. The simplest model must have $h_{(1,1)}=3$ since on the
heterotic side, compactified on $K_3\times S_1$, one gets at least
three vector fields: $g_{\mu 6}$, $ b_{\mu 6}$ and the antisymmetric
tensor $b_{\mu\nu}$ which is dual to a vector in
$D=5$. One of the vector bosons  corresponds to the graviphoton of the
gravitational supermultiplet while the remaining  two form vector
multiplets whose (real) scalar components are the 5-D dilaton $\phi$
and the radius field $R$ associated with  $S_1$. A possible way to reduce the
number of vectors is to consider a 5-D theory with a fixed radius $R_0$.
In that case only one combination of $g_{\mu 6}$, $b_{\mu 6}$ appears which
corresponds to the graviphoton, and the theory is related to $h_{(1,1)}=2$
Calabi Yau threefold. In general if a gauge group of rank $r$ is added so that
$h_{(1,1)}=2+r$, the scalar components of vector multiplets parameterize the
space
\be
{\cal M}=O(1,1)\times {O(1,r)\over O(r)}.\label{cos}
\en

It is well known that, in five dimensions, the
$(h_{(1,1)}{-}1)$-dimensional space $\cal M$ of scalar components of
$N=2$ abelian vector multiplets coupled to supergravity can be
regarded as a  hypersurface of a $h_{(1,1)}$-dimensional
manifold whose coordinates $X$ are in correspondence with the vector
bosons, including the graviphoton \cite{g}. The equation of the
hypersurface is ${\cal V}(X)=1$, where ${\cal V}$ is a homogeneous
cubic polynomial in $X$'s. If ${\cal V}$ is factorizable, \ie\ of
the form ${\cal V}=sQ(t)$, where $Q$ is a quadratic form in $t$'s,
then $\cal M$ is given precisely by (\ref{cos}).

It is also known that the  function ${\cal V}(X)$ describing the
vector multiplet sector of 11-D supergravity compactified on a
Calabi-Yau threefold (CY) is given by the CY intersection form
\cite{cfg,ht,cf}. If we consider a threefold obtained by a $K_3$ fibration
\cite{klm}, then
\be
{\cal V}=sQ(t)+C(t)\ , \label{nu}
\en
where $Q(t)$ is quadratic and $C(t)$ is cubic. {}From the heterotic
point of view, identifying $s$ as the inverse of the string loop
expansion parameter ($s\sim 1/g^2$), we see that the first term
corresponds to a tree-level contribution and gives the desired scalar
manifold (\ref{cos}). Hence $K_3$ fibrations provide good candidates
for CY duals of the heterotic theory. The cubic form $C(t)$ should be
obtained by a one-loop calculation in the heterotic theory,
reproducing all remaining CY intersection numbers. In this paper we
show that this is indeed the case.

The paper is organized as follows. In section 2, we discuss the basic
features of 5-D theories obtained by compactifying 11-D supergravity
on Calabi-Yau threefolds. For CY manifolds which are $K_3$ fibrations,
we perform a duality transformation which
brings the lagrangian to a form that can be compared with the 5-D
heterotic superstring, with one of the vector bosons replaced by the
antisymmetric tensor field. In section 3, we determine the one-loop
effective action by computing appropriate heterotic superstring
amplitudes in a rank 2+1 model. Our results agree with the known form of the
4-D
$N=2$ prepotential of the heterotic theory compactified on $K_3\times T_2$
\cite{afgnt,dw}, in the decompactification limit in which one of the
torus radii goes to infinity. We argue that the one-loop results are
exact, at least in some finite region of the moduli space. In section
4, we discuss duality between the heterotic theory and 11-D
supergravity compactified on Calabi-Yau threefolds, providing
a dual description of the enhanced gauge symmetry points which unlike
in four dimensions remain present in the full quantum theory.\footnote{Enhanced
gauge symmetries on CY manifolds may also appear in some cases in $D=4$ as
recently discussed in refs.\cite{asp1,bvs}.} In section 5, we exhibit the
existence of additional massless states which appear in the strong coupling
regime of the heterotic theory.

\section{11-D Supergravity Compactified to $D=5$ on Calabi-Yau
Threefolds}

In this section we will recall the basics of the low-energy theory of
11-D supergravity compactified on a generic Calabi-Yau threefold with
Hodge numbers $h_{(1,1)}$, $h_{(2,1)}$  and the intersection numbers
$C_{\Lambda\Sigma\Delta}$ ($\Lambda,\Sigma,\Delta=1,\dots,h_{(1,1)}$)
\cite{cf}. The bosonic fields of 11-D theory are the elfbein
$e_{\hat{\mu} \hat{a}}$ and the  three-form gauge field
${\cal A}_{\hat{\mu}\hat{\nu}\hat{\rho}}$, with all indices running
over $1,\dots,11$. It is convenient to split these indices as
$\hat{\mu}=(\mu,i,\bar{\jmath})$,
$\mu=1,\dots,5$ and $i,\bar{\jmath}=1,2,3$. The $h_{(1,1)}$ moduli
split then into $h_{(1,1)}-1$  moduli with unit volume ($\det
g_{i\bar{\jmath}}=1$) and the volume modulus $\det g_{i\bar{\jmath}}$.
The massless spectrum contains $h_{(1,1)}-1$ vector multiplets with
real scalar components given by the moduli at unit volume. The vector
bosons, including the graviphoton, are the $h_{(1,1)}$ one-forms (on
space-time) ${\cal A}_{\mu i\bar{\jmath}}$. There is one universal
hypermultiplet with the scalar components $(\det g_{i\bar{\jmath}},
{\cal A}_{\mu\nu\rho},{\cal A}_{ijk}=\epsilon_{ijk}a)$. There are also
$h_{(2,1)}$  additional hypermultiplets whose scalar components are
given by the complex scalar pairs $(g_{i{\bar\jmath}}, {\cal
A}_{ij\bar{k}})$. For our purposes, the most important fact is the
absence from the spectrum of a scalar corresponding to a two-index
antisymmetric tensor field with both internal indices. The absence
of such a field implies that there are no non-perturbative instanton
corrections to the low-energy effective action describing the vector multiplet
sector of the theory \cite{bbs}.

The effective $N=2$ supersymmetric lagrangian describing vector
multiplets  coupled to supergravity is completely determined by one
function ${\cal V}(X)$, a homogeneous cubic polynomial of the vector
coordinates $X^{\Lambda}$ \cite{g}. As already mentioned before, in the
case of a theory obtained by compactifying 11-D supergravity on a
Calabi-Yau threefold, this function is given by the intersection form
\cite{cf}:\footnote{This corresponds to the ``very special geometry''
of ref.\cite{dwp}.}
\be
{\cal V}=\frac{1}{6}C_{\Lambda\Sigma\Delta} X^{\Lambda}
X^{\Sigma}X^{\Delta}\label{cxxx}
\en
The bosonic part of the lagrangian is given then by \cite{g}
\be
{\cal L}_{\makebox{\small{b}}}=-{\frac{1}{2}}{\cal R}
-\frac{1}{2}g_{xy}\partial\phi^x\partial\phi^y
-\frac{1}{4}G_{\Lambda\Sigma} F^{\Lambda}F^{\Sigma}+
\frac{1}{48}C_{\Lambda\Sigma\Delta}\epsilon F^{\Lambda}F^{\Sigma} A^{\Delta}\ ,
\label{lbos}
\en
where $\cal R$ is the Ricci scalar
and in the last term the space-time indices are contracted by using
the completely antisymmetric 5-D $\epsilon$-tensor. The scalars parameterize
the hypersurface ${\cal V}(X)=1$, and their metric is related to the vector
metric by
\be
g_{xy}=G_{\Lambda\Sigma}\left.\partial_{\phi_x}X^{\Lambda}
\partial_{\phi_y}X^{\Sigma}\right|_{{\cal V}=1}.\label{gs}
\en
{}Finally, the vector metric
\be
G_{\Lambda\Sigma}=-\frac{1}{2}
\left.\partial_{\Lambda}\partial_{\Sigma}\ln{\cal V} \right|_{{\cal V}=1}.
\label{gv}
\en

{}For a generic Calabi-Yau manifold there is no preferred vector field,
however  in the case of manifolds obtained by $K_3$ fibrations,
the form (\ref{nu}) of the function $\cal V$ singles out $s$.
We will dualize the gauge vector field $A^s_{\mu}$
into an antisymmetric tensor and identify it with the $b_{\mu\nu}$
field of the dual heterotic string theory.  The scalar manifold
will be parameterized by $t_k$ with $k=1,\dots,h_{(1,1)}-1$; $s$
will be eliminated by using the constraint
\be
{\cal V} =1 ~\Rightarrow~    s = \frac{1-C(t)}{Q(t)}\ .\label{co}
\en
Using eq.(\ref{gv}) with the constraint (\ref{co}) one obtains
\begin{eqnarray}
G_{ss}&=&\frac{1}{2}Q^2\nonumber\\
G_{sk}&=&\frac{1}{2}(Q\partial_kC-C\partial_kQ)\label{metr}\\
G_{kl}&=& \frac{1-C}{2}(-\partial_k\partial_l\ln Q +\partial_kC\partial_l\ln Q
+\partial_kC\partial_l\ln Q-C\partial_k\ln Q\partial_l\ln Q)\nonumber\\ &
&\qquad
+\,\frac{1}{2}(\partial_kC\partial_lC-\partial_k\partial_lC)\nonumber
\end{eqnarray}

The vector field  $A^s_{\mu}$ can be dualized by introducing
a Lagrange multiplier term
\be
{\cal L}_{LM}=\frac{1}{48}
\epsilon^{\mu\nu\lambda\rho\sigma}F^s_{\mu\nu}H_{\lambda\rho\sigma}\label{lm}
\en
where $H_{\lambda\rho\sigma}=
\partial_{\lambda}b_{\rho\sigma} +\partial_{\sigma}b_{\lambda\rho}
+\partial_{\rho}b_{\sigma\lambda} $
is the field strength of the antisymmetric tensor field.
The antisymmetric tensor field equations impose the Bianchi identity on
$F_s$.  The duality transformation is performed by eliminating
$F_s$ with the use of its equation of motion:
\be
2G_{ss}F_s+2G_{sk}F_k-\frac{1}{4}Q_{kl}\,\epsilon F_kA_l
-\frac{1}{12}\epsilon H=0\ ,\label{due}
\en
where $Q_{kl}\equiv\partial_k\partial_l Q$.
After substituting $F_s$ into eq.(\ref{lbos}) the bosonic part of
the lagrangian becomes
\begin{eqnarray}
{\cal L}_{\makebox{\small{b}}}&=&-{\frac{1}{2}}{\cal R}
-\frac{1}{2}g_{xy}\partial\phi^x\partial\phi^y +{1\over 192G_{ss}}H^2
-\frac{1}{4}(G_{kl}-{G_{sk}G_{sl}\over G_{ss}})F_kF_l
+{1\over 32G_{ss}}Q_{kl}HF_kA_l\nonumber\\
& &-{G_{sk}\over 48G_{ss}}\epsilon HF_k
+{1\over 48}(C_{klm}-{3G_{sk}\over G_{ss}}Q_{lm})
\epsilon F_kF_lA_m +{3\over 64G_{ss}}(Q_{kl}F_kA_l)^2
\label{lagr}
\end{eqnarray}

As mentioned in the introduction, in the heterotic theory $1/s$ plays the
role of the 5-D string coupling constant.  The tree level metric of the
gauge fields is then given by $G_{kl}=-{1\over 2}\partial_k\partial_l\ln Q$
and  the only non-vanishing ``Yukawa" couplings involve the antisymmetric
tensor with two gauge fields. The couplings between three gauge bosons,
as well as the mixing of gauge bosons with the antisymmetric tensor
will appear at the one loop level.

\section{Heterotic Superstring Compactified to $D=5$ on $K_3\times S_1$}

$N=2$ supersymmetric 5-D theory obtained by compactifying heterotic
superstring on $K_3\times S_1$ contains a gauge group with the rank ranging
from 2 to 2+21, depending on details of compactification. The 2 gauge
bosons are universal: the graviphoton and the vector dual to the
antisymmetric tensor $b_{\mu\nu}$. Here we shall consider a rank 3 example
with one additional $U(1)$ gauge boson associated to $S_1$. It can be
constructed by following the lines of \cite{kv} and its further $S_1$
compactification to four  dimensions yields the rank 4 model which on the type
II side is described by $X_{24}(1,1,2,8,12)$ CY compactification
\cite{hk,klm}. This model contains also 244 massless neutral
hypermultiplets. We will derive the exact effective action describing the
interactions of vector multiplets at the two-derivative level.

The vector moduli space of the above rank 3 model contains 2 real
scalars, the dilaton $\phi$ whose expectation value determines the 5-D
string coupling constant and the radius field $R$ whose expectation value
determines the radius of the circle $S_1$. At a generic point of this
moduli space, the gauge group is $U(1)^3$, with the gauge bosons
$b_{\mu\nu}$, $g_{\mu 6}$ and $b_{\mu 6}$, and there are no massless
charged states. The massive Kaluza-Klein excitations and winding  modes
associated to $S_1$ are charged with respect to $g_{\mu 6}$ and  $b_{\mu
6}$, but they are neutral with respect to $b_{\mu\nu}$. Their left and
right momenta belong to a $O(1,1)$ lattice,
\be
p_{R,L}={1\over {\sqrt 2}}({m\over R}\pm nR)\ ,
\label{pLpR}
\en
with integer $m$ and $n$, therefore their masses depend on the radius $R$.
At $R=1$ two additional massless vector multiplets appear with $m=n=\pm 1$, so
that $p_L=0$ and $p_R=\pm{\sqrt 2}$, and the $U(1)$ factor corresponding to
the combination $A_{\mu}{\sim} g_{\mu 6}{+}b_{\mu 6}$ gets enhanced to
$SU(2)$.

We will now derive the effective action by computing the appropriate
superstring amplitudes. The string vertices for the gauge fields
$A$ and $B$ corresponding to the right- and left-moving
combinations $g_{\mu 6}{\pm}b_{\mu 6}$, in the ($-$1)-ghost picture,
are:
\begin{eqnarray}
V^\mu_A(p,z) &=& :\psi^\mu\dbar X_6 e^{ip\cdot X}:\label{VA}\\
V^\mu_B(p,z) &=& :\psi_6\dbar X^\mu e^{ip\cdot X}:\label{VB}
\end{eqnarray}
where $X^\mu$ and $X_6$ are the space-time and $S_1$ coordinates,
respectively,
while $\psi^\mu$ and $\psi_6$ are their world-sheet fermionic
superpartners. The vertices for the graviton, dilaton ($V_{\phi}$) and
antisymmetric tensor  are respectively the symmetric-traceless, trace
and antisymmetric parts of
\be
V^{\mu\nu}(p,z) = :\psi^{\mu}\dbar X^{\nu} e^{ip\cdot X}:
\label{Vgr}
\en
{}Finally, the vertex for the radius is:
\be
V_R(p,z) = :\psi_6\dbar X_6 e^{ip\cdot X}:
\label{VR}
\en

As explained in the introduction, the general form of the function $\cal V$
which determines the low energy effective action on the heterotic side is
\be
{\cal V}=sQ(A,B)+C(A,B)
\en
where the first term represents the tree level contribution depending
on the quadratic form $Q(A,B)$ and the one loop correction is described
by the cubic function $C(A,B)$.  $Q(A,B)$ can be obtained from
the tree-level three-point amplitudes involving one antisymmetric
tensor and two gauge fields. It is easy to see that the only non-vanishing
amplitudes are  $\lv bAA\rv$ and $\lv bBB\rv$, giving
\be
Q(A,B)=A^2-B^2 \label{Q}
\en
It follows then from eq.(\ref{lagr}) that the gauge kinetic terms are not
diagonal which can be confirmed by calculating the amplitude $\lv R AB\rv$.
Hence it is convenient to
diagonalize the gauge kinetic terms by changing the vector field
basis to
\be
t_1=A+B~~~~~~~~~~t_2=B-A\ ,\label{rel}
\en
which is equivalent to going back
to  $g_{\mu 6}(=A_{1\mu})$ and $ b_{\mu 6}(=-A_{2\mu})$.
The scalar field surface ${\cal V}=1$ can be parameterized by
\be
R=\left({t_1\over t_2}\right)^{1/2}~~~~~~~~~~~~~\phi={2\pi\over t_1t_2}\ .
\label{rels}
\en
In this field basis, $Q=t_1t_2$, and the the tree-level bosonic part of the
lagrangian (\ref{lagr}) becomes
\be
{\cal L}_{\makebox{\small{b}}}^{\makebox{\it tree}}=-{\frac{1}{2}}{\cal R}
-{\phi\over 16\pi} (\frac{1}{R^2}F_1^2+R^2F_2^2)
+{\phi^2\over 16\pi^2}({1\over 24}H^2+\frac{1}{2}HF_1A_2)
-\frac{1}{2}\frac{(\partial R)^2}{R^2} -\frac{3}{8}
\frac{(\partial \phi)^2}{\phi^2}
\label{lb}\en
The assignment of  vertices $V_{\phi}$ and $V_R$ to the scalars $\phi$
and $R$ can be checked  by computing three-point amplitudes involving
one of these  scalars and two gauge bosons. Indeed, using the vertices
(\ref{VA}-\ref{VR}) one can show
that the only non-vanishing amplitudes\footnote{As usual, the space-time
momenta
have to be complexified in order to avoid kinematical constraints which  make
these amplitudes zero on-shell.} of this type are $\lv\phi AA\rv$,
 $\lv\phi BB\rv$, $\lv R AB\rv$ and  $\lv\phi bb\rv$, in agreement with
the lagrangian (\ref{lb}) and the relations (\ref{rel}-\ref{rels}).

The one-loop function
\be
C=a_1t_1^3 +a_2t_2^3+b_1t_1^2t_2+b_2t_2^2t_1 \label{c}
\en
is parameterized by four constants $a_{1,2}$ and $b_{1,2}$. In fact,
the physical amplitudes depend on $a_{1,2}$ only, since
the last two terms can be eliminated by shifting the ``dilaton'' $s$,
\be
s\rightarrow s-b_1t_1-b_2t_2\ , \label{sshift}
\en
which corresponds to a perturbative symmetry of the heterotic theory.
In order to extract $a_{1,2}$ it is sufficient to consider the ``Yukawa''
couplings between three gauge bosons. The relevant interaction
terms obtained from eq.(\ref{lagr}) have the form
\be
{\cal L}_{Y}^{\makebox{\it 1-loop}}=
\frac{1}{48}a_1\epsilon F_1F_1A_1+
\frac{1}{8}(-2a_1R^2+{a_2\over R^4})\epsilon F_1F_1A_2
+ (1\leftrightarrow 2, R\leftrightarrow {1\over R})\ .
\label{ly}
\en
Using the above expression, it is straightforward to express
all 3-gauge boson amplitudes in terms of the unknown constants
$a_{1,2}$. In order to make contact with the superstring computation,
it is convenient to go back to the $A,B$ basis (\ref{rel}) of the
vertex operators. One finds that the amplitudes $\lv BBB\rv$
and $\lv BBA\rv$ vanish while
\begin{eqnarray}
\lv A_{\mu}(p_1)A_{\nu}(p_2)B_{\lambda}(p_3)\rv
&=&\frac{1}{2}\,\epsilon_{\mu\nu\lambda\rho\sigma}p_1^{\rho}p_2^{\sigma}
(a_1R^3+{a_2\over R^3})\label{amps1}\\
\lv A_{\mu}(p_1)A_{\nu}(p_2)A_{\lambda}(p_3)\rv
&=&\frac{3}{2}\,\epsilon_{\mu\nu\lambda\rho\sigma}p_1^{\rho}p_2^{\sigma}
(a_1R^3-{a_2\over R^3})\label{amps2}
\end{eqnarray}

The above one-loop amplitudes receive contribution from the odd spin structure
only. There is one gauge boson vertex in the ($-$1)-ghost picture and the
other two in the 0-picture which are obtained from (\ref{VA},\ref{VB}) by
replacing $\psi^\mu$ ($\psi_6$) by $\partial X^\mu{+}ip\cdot \psi\psi^\mu$
($\partial X_6{+}ip\cdot \psi\psi_6$). In addition there is a world-sheet
supercurrent insertion,
\be
T_F=:\psi^\mu\partial X_{\mu}+\psi_6\partial X_6:+T_F^{\makebox{\small{int}}}
\label{TF}
\en
where $T_F^{\makebox{\small{int}}}$ represents the internal $K_3$ part,
as well as the
ghost contribution. In the odd spin structure, six fermionic zero-modes
must be saturated to yield a non-zero result. It follows that the amplitudes
$\lv BBB\rv$ and $\lv BBA\rv$ vanish as expected, while in the other two the
saturation of the fermionic zero-modes gives rise to the $\epsilon$-tensor
as in
eqs.(\ref{amps1},\ref{amps2}). One obtains:
\begin{eqnarray}
\lv A_{\mu}(p_1)A_{\nu}(p_2)B_{\lambda}(p_3)\rv
&=&\epsilon_{\mu\nu\alpha\rho\sigma}p_1^{\rho}p_2^{\sigma}
\int_{\Gamma}{d^2\tau\over\tau_2}\prod_{i=1}^3\int[d^2z_i]\nonumber\\
& &\quad
\lv\dbar X_6(\zbar_1)\dbar X_6(\zbar_2)\dbar X_\lambda(\zbar_3) \partial
X^\alpha(0)\rv_{\makebox{\small{odd}}}\label{AAB}\\
\lv A_{\mu}(p_1)A_{\nu}(p_2)A_{\lambda}(p_3)\rv
&=&\epsilon_{\mu\nu\lambda\rho\sigma}p_1^{\rho}p_2^{\sigma}
\int_{\Gamma}{d^2\tau\over\tau_2}\int\prod_{i=1}^3[d^2z_i]\nonumber\\
& &\quad
\lv\dbar X_6(\zbar_1)\dbar X_6(\zbar_2)\dbar X_6(\zbar_3) \partial X_6(0)
\rv_{\makebox{\small{odd}}}\label{AAA}
\end{eqnarray}
where $\tau = \tau_1 + i\tau_2$ is the Teichm\"uller parameter of the
world-sheet torus and ${\Gamma}$ its fundamental domain.

The contraction of
$\dbar X_{\lambda}$ with $\partial X^{\alpha}$ in eq.(\ref{AAB})
gives $\lv\dbar X_{\lambda}(\zbar_3) \partial X^{\alpha}(0)\rv =
-\delta_{\lambda}^{\alpha}{\pi}/{4\tau_2}$. The two $\dbar X_6$
insertions are replaced by their zero modes since their contractions yield
total derivatives which vanish upon $z_i$ integration. Similarly in
eq.(\ref{AAA}), $\partial X_6$ and all $\dbar X_6$'s are replaced by their zero
modes. After performing the $z_i$ integrations and taking into account the
left-moving part of all determinants, we find:
\begin{eqnarray}
\lv A_{\mu}(p_1)A_{\nu}(p_2)B_{\lambda}(p_3)\rv
&=&\frac{1}{2}\epsilon_{\mu\nu\lambda\rho\sigma}
p_1^{\rho}p_2^{\sigma}\, {\cal I} \label{AAB1}\\
\lv A_{\mu}(p_1)A_{\nu}(p_2)A_{\lambda}(p_3)\rv
&=&\frac{3}{2}\epsilon_{\mu\nu\lambda\rho\sigma}p_1^{\rho}p_2^{\sigma}\,
R\partial_R{\cal I}
\label{AAA1}
\end{eqnarray}
where
\be
{\cal I}=\frac{1}{16}\int_{\Gamma}{d^2\tau\over\tau_2^2}
\partial_{\taubar}(\tau_2^{1/2}Z){\bar F}(\taubar)
\label{I}
\en
with $Z$ being the $S_1$ partition function,
\be
Z = \sum_{p_L,p_R} e^{i\pi\tau p_L^2}
e^{-i\pi\taubar p_R^2}\ .
\label{Z}
\en
${\bar F}(\taubar)$ is an $R$-independent antimeromorphic form of weight
$-2$ in $\taubar$ with a simple pole at infinity due to the tachyon of the
bosonic sector \cite{afgnt}, ${\bar F}\sim
\frac{i}{\pi^2}e^{-2i\pi\taubar}$. In deriving eq.(\ref{AAA1}) we used the
explicit form of lattice momenta (\ref{pLpR}).

To evaluate the integral ${\cal I}$ we start from the identity:
\be
[(R\partial_R)^2-1]Z=16\tau_2^{3/2}
\partial_\tau\partial_{\taubar}(\tau_2^{1/2}Z)
\label{relZ}
\en
which implies the following differential equation:
\be
[(R\partial_R)^2-9]{\cal I}=\int_{\Gamma}d^2\tau{\bar F}
(\taubar)\partial_\tau\{ {1\over\tau_2^2}
\partial_{\taubar}[\tau_2^2\partial_{\taubar}(\tau_2^{1/2}Z)]\}
\label{relI}
\en
The r.h.s.\ being a total derivative with respect to $\tau$ vanishes away
from the enhanced symmetric point $R=1$. At $R=1$, the surface term gives
rise to a $\delta$-function due to singularities associated to the
additional massless particles which enhance the gauge symmetry to
$SU(2)\times U(1)^2$. Expanding $p_L$, $p_R$ around $R=1$ for these states,
it is easy to show that the surface term becomes proportional to:
$$
\lim_{\tau_2\rightarrow\infty} \tau_2^{1/2} e^{-\pi\tau_2(R-{1\over R})^2}
=\frac{1}{2}\delta(R-1)\ .
$$
In terms of the ``time'' variable $\ln R$, eq.(\ref{relI}) becomes the
one-dimensional propagator equation with the mass squared $-9$. Its general
solution is
\be {\cal I}=\frac{1}{3} e^{-3|\ln R|}+\alpha(R^3+{1\over R^3})\ , \label{sol}
\en
where $\alpha$ is an arbitrary constant depending on the boundary conditions.
Since at $R\rightarrow\infty$ ${\cal I}\rightarrow 0$ as ${\cal O}(R^{-3})$,
\cf\ eq.(\ref{I}), $\alpha=0$, so that
\be {\cal I}=\frac{1}{3}[\theta (R-1){1\over R^3}+\theta (1-R)R^3]\ .
\label{isol}
\en
By comparing eqs.(\ref{AAB1},\ref{AAA1}) with (\ref{amps1},\ref{amps2})
we find
\be
a_1=\frac{1}{3}\theta (1-R)\, ,\qquad\qquad a_2=\frac{1}{3}\theta (R-1)\ .
\label{as}\en

This result can also be checked by studying the mixing between
the antisymmetric tensor and the gauge bosons.
The relevant lagrangian term, \cf\ eq.(\ref{lagr}),  is
\be
{\cal L}^{\makebox{\it{1-loop}}}_{H F}=\frac{1}{48}
(-2a_1R^2+{a_2\over R^4})\epsilon HF_1+
 (1\leftrightarrow 2, R\leftrightarrow {1\over R})\ .\label{hf}
\en
These interactions generate in particular the amplitudes
involving one antisymmetric tensor field, one of the gauge bosons and
one scalar. Going back to the $A,B$ basis, it is easy to see that the only
non-vanishing amplitudes involve the gauge field $A$ and the radius
scalar $R$:
\be
\lv A_\mu(p_1) b_{\nu\lambda}(p_2) R(p_3)\rv =
\frac{1}{2}\,\epsilon_{\mu\nu\lambda\rho\sigma}p_1^{\rho}p_2^{\sigma}
(a_1R^3+{a_2\over R^3})
\label{AHR}
\en
The superstring computation of such  amplitudes proceeds
in a similar way as in the previous case.  Only the odd spin structure
contributes, and the zero-mode counting argument shows that
(\ref{AHR}) is indeed the only non-vanishing amplitude.
One finds
\begin{eqnarray}
\lv A_\mu(p_1) b_{\nu\lambda}(p_2) R(p_3)\rv &=&
\epsilon_{\mu\nu\alpha\rho\sigma}p_1^{\rho}p_2^{\sigma}
\int_{\Gamma}{d^2\tau\over\tau_2}\prod_{i=1}^3\int[d^2z_i]\nonumber\\
& &\quad
\lv\dbar X_6(\zbar_1)\dbar X_\lambda(\zbar_2)\dbar X_6(\zbar_3)) \partial
X^\alpha(0)\rv_{\makebox{\small{odd}}}.\label{AHR1}
\end{eqnarray}
This correlation function can be evaluated as (\ref{AAB}) with the result
\be
\lv A_\mu(p_1) b_{\nu\lambda}(p_2) R(p_3)\rv =\frac{1}{2}
\epsilon_{\mu\nu\lambda\rho\sigma}p_1^{\rho}p_2^{\sigma}\, {\cal I}
\en
which becomes compatible with the field-theoretical expression (\ref{AHR})
after using eqs.(\ref{isol}) and (\ref{as}).

To summarize, the result of the heterotic superstring computation
is the function
\be
{\cal V}=st_1t_2+ \frac{1}{3}\theta(t_2-t_1)t_1^3 +\frac{1}{3}
 \theta(t_1-t_2)t_2^3 \label{fin} \en
The discontinuity at $t_1=t_2$ ($R=1$) is due to the appearance
of massless particles at the enhanced symmetry point. It is analogous
to the logarithmic singularity of the prepotential in $D=4$,
however the infrared behavior of five-dimensional theory is
different than in four dimensions.

In fact, eq.(\ref{fin}) can be derived by studying the 4-D theory obtained
by a compactification on $K_3\times T_2$ in the limit when
one of the $T_2$ radii goes to infinity.
In the rank 4 ($STU$) model of refs.\cite{kv,klm}, the moduli space of
$T_2$ is characterized by two complex moduli $T$ and $U$ which parameterize
the metric and the antisymmetric tensor $b_{56}$. Taking $b_{56}=0$
and a diagonal $T_2$ metric, with $g_{55}=R^2_5/2\rightarrow\infty$ and
$g_{66}=R^2/2$ we have:
\be
T=iR_5{R\over\sqrt{2\phi}}\qquad U=i{R_5 \over R\sqrt{2\phi}} \qquad
S=2iR_5\phi\ .\label{stu}
\en
These equations follow from the usual relations between 5-D and 4-D
scalars \cite{g} upon the identification
$t_1\rightarrow T$, $t_2\rightarrow U$ and $s\rightarrow S$ as dictated
by the form of the lagrangian (\ref{lb}). Eqs.(\ref{stu}) imply
that all three moduli go to infinity in the decompactification limit.
The order of the these limits correspond to different domains
of the ($R,\phi$) space. The 5-D perturbative region corresponds to
\be
(2\phi)^{-3/2}<R<(2\phi)^{3/2}
\label{pert}\ ,
\en
which implies that the weak coupling limit $S\rightarrow
i\infty$ has to be taken first. The tree-level ``Yukawa" coupling
$f_{STU}=1$ coincides with $C_{s12}$ in five dimensions upon the
identification $s=S$, $t_1=T$, $t_2=U$. Then,
using the one loop result \cite{afgnt,dw}\footnote{We make a factor 1/2
adjustment to the result \cite{afgnt} taking into account different
normalization of $S$.}
\be
f_{TTT}=-{i\over\pi}\frac{j_T(T)}{j(T)-j(U)}
\left\{\frac{j(U)}{j(T)}\right\}
\left\{\frac{j_T(T)}{j_U(U)}\right\}
\left\{\frac{j(U)-j(i)}{j(T)-j(i)}\right\}
\label{fTTT}
\en
we find that in the $R_5\rightarrow\infty$ limit
\be
f_{TTT}\rightarrow
{2\over 1-e^{2\pi(R-1/R)R_5/\sqrt{2\phi}}}\rightarrow
2\theta(1-R)=6a_1=C_{111}\ .
\label{a1}
\en
Similarly,
\be
f_{UUU}\rightarrow 2\theta(R-1)=6a_2=C_{222}\ .
\label{a2}
\en
The above expressions agree with the result (\ref{as})
obtained by means of a direct computation in $D=5$.

In four dimensions, the infinite moduli limit taken in the manner
described above forces the effective theory into the perturbative regime,
suppressing non-perturbative effects. This occurs, however, not only in
the limit of asymptotically small 5-D coupling, but in the finite
interval (\ref{pert}). Thus, the heterotic result (\ref{fin}) is exact
in this region of the parameter space.  $SU(2)$ gauge group remains
unbroken at $R=1$ at the non-perturbative level.

A similar analysis is even simpler for models of rank 2 which upon
compactification to $D=4$ become equivalent to type II superstring compactified
in CY manifolds with $h_{(1,1)}=3$, like $X_{12}(1,1,2,2,6)$ and
$X_8(1,1,2,2,2)$ \cite{kv,ap}. In this case there is no enhanced gauge symmetry
point and the heterotic tree-level action with ${\cal V}\sim st^2$ does not
receive loop corrections in the weak coupling region $t<1$.

\section{Heterotic Superstring, 11-D Supergravity and $p$-branes}

In this section we will discuss some aspects of duality between $N=2$
supersymmetric heterotic superstring in five dimensions and 11-D
supergravity theory compactified on a Calabi-Yau threefold.  The latter
contains BPS states obtained by wrapping two- and five-branes on even CY
cycles.  It has been argued before that 11-D supergravity describes the
strong coupling limit of 10-D type IIA superstring theory \cite{w}. In order to
see a similar connection in $D=5$, it is convenient to consider first further
compactification from $D=5$ to $D=4$ on $S_1$.
Then using duality between heterotic theory on $K_3\times S_1\times S_1$
and type II on Calabi-Yau, the 5-D decompactification limit $R_5\rightarrow
\infty$ corresponds to the large volume (complex structure) limit of the
CY manifold on the type IIA (IIB) side, see eq.(\ref{stu}).

In the previous section we have shown that the effective
action describing the heterotic side exhibits exact singularities
due to enhanced gauge symmetries. These singularities,
as viewed from the type II side, provide a strong evidence
for the presence of enhanced gauge symmetry points on
CY threefolds obtained by $K_3$ fibrations, in the large complex structure
limit.
This is similar to the case of $K_3$
twofolds where the presence of enhanced symmetries is
related to duality in seven dimensions, between 11-D theory compactified on
$K_3$ and heterotic superstring theory compactified on $T_3$
\cite{w,branes,asp2}.

Let us first consider the central charge formula in five dimensions,
for a generic supergravity theory. From the supersymmetry
algebra it follows that the central charge is
\be
Z_e=\sum_{\Lambda}t^{\Lambda}e_{\Lambda}\ , \label{ze}
\en
where $t^{\Lambda}=(s,t^i)$ are the $D=5$  special coordinates and
$e_{\Lambda}$ are the electric charges. The dual formula
for the ``magnetic'' charges (string-like objects) is
\be
Z_m=\sum_{\Lambda}t_{\Lambda}m^{\Lambda}\ , \label{zm}
\en
where $\sum_{\Lambda}t^{\Lambda}t_{\Lambda} =1$, therefore
\be
t_{\Lambda}=C_{\Lambda\Sigma\Delta}t^{\Sigma}t^{\Delta}\ . \label{tt}
\en
{}From the heterotic point of view  $e_i$ correspond to the usual
(perturbative) electric charges of Kaluza-Klein excitations and winding modes,
$m^s$ is the charge of the fundamental string, while $m^i$ and $e_s$ arise from
10-D solitonic five-branes wrapping around $K_3$ and $K_3\times S_1$,
respectively \cite{s1,dm,hm}. In the dual 11-D supergravity theory, these
states
originate from two- and five-brane solitons which wrap even cycles in the
Calabi-Yau space \cite{cf,bbs}:
\be
e_{\Lambda}=\int_{{\cal C}^{4\Lambda}\times S_3}G_7\ ,~ \quad\qquad
m^{\Lambda}=\int_{{\cal C}_{2}^{\Lambda}\times S_2}F_4 \ ,\label{emchar}
\en
where $F_4$ is the field strength of the three-index antisymmetric tensor field
and $G_7=\delta{\cal L}/\delta F_4$ is its dual; ${\cal C}_2^{\Lambda}$ and
${\cal C}_{4\Lambda}$, $\Lambda=1,\dots,h_{(1,1)}$,
are two- and four-cycles  in the CY space, respectively, while $S_2$ and $S_3$
are two- and three-dimensional spheres in 5-D spacetime.

In the case under discussion with two vector multiplets and
${\cal V}=\frac{1}{6}C_{\Lambda\Sigma\Delta} t^{\Lambda}
t^{\Sigma}t^{\Delta}=sQ(t)+C(t)$, we have
\be
Z_e  ~=~ se_s+t^1e_1+t^2e_2 ~=~  { 1-C(t)\over Q(t)}e_s+t^1e_1+t^2e_2\ .
\label{z1}\en
{}For $C(t)=0$, this formula gives
\be
Z_e={1\over g_5^2}e_s+g_5(Re_1+{1\over R}e_2)\ ,\label{z2}
\en
where $g_5^2\equiv 2\pi/\phi$.
Note that eq.(\ref{z2}) reproduces the $O(1,1)$ Narain lattice
($e_s=0$) and also the Witten formula ($e_1=e_2=0, e_s\neq 0$) \cite{w}
for the non-perturbative states which are electrically charged with respect to
the $b_{\mu\nu}$ field. In the presence of one-loop corrections $C(t)$
calculated in the previous section, the central charge becomes:
\be
Z_e=({1\over g_5^2}-\frac{1}{3}g_5R^3)e_s+g_5(Re_1+{1\over R}e_2)\ ,\label{z3}
\en
for $R<1$ and a similar expression with $R\rightarrow 1/R$ for $R>1$.

Eq.(\ref{z3}) raises the question about existence
of massless states with vanishing $Z_e$ (and/or $Z_m$).
On the heterotic side, we do certainly have enhanced gauge symmetry
at $R=1$ and possibly also at some other non-perturbative point or
line with $g_5$ related to $R$. What is the interpretation of
these points on the CY side? Considering 5-D CY theory as
a decompactification limit of the 4-D theory corresponds to
taking the limit $S\rightarrow i\infty, ~T\rightarrow i\infty,
{}~U\rightarrow i\infty$
while keeping all ratios fixed, see eq.(\ref{stu}). If we send
$S\rightarrow i\infty$ first, we obtain the Yukawa coupling of ref.\cite{hk} in
agreement with the perturbative heterotic computation \cite{afgnt}.
The result (\ref{a1},\ref{a2}) exhibits a discontinuity due
to the existence of an enhanced symmetry point at $T=U$.
The other enhanced symmetry points, associated to $SU(3)$ and $SO(4)$
gauge groups, disappear for large $T, U$,  however the $SU(2)$ gauge group
remains intact at $T=U$ ($R=1$).

In the decompactification limit, the mass of a state as measured in
$D=5$ is related to its original 4-D mass in the following way:
\be M_5^2=\lim_{R_5\rightarrow\infty}R_5\, M^2_4(R_5) \label{m5}
\en
{}Furthermore, by comparing 5-D and 4-D theories it is easy to show
that the respective moduli are related by \cite{g}:
\be
T^{\Lambda}=t^{\Lambda}R_5\ , \label{t5}
\en
where $T^{\Lambda}$ and $t^{\Lambda}$ are the 4-D and 5-D moduli,
respectively, so that $C_{\Lambda\Sigma\Delta} T^{\Lambda}
T^{\Sigma}T^{\Delta}=R_5^3$, in agreement with the standard supergravity
result.  Starting from the 4-D BPS mass formula \cite{bps}
specified to the case of  large $S$, $T$ and $U$ moduli,
with the K\"ahler potential
\be
K(S,T,U)\sim -\ln{\cal V}\sim -3\ln R_5\ ,\label{kal}
\en
one obtains (in the absence of magnetic charges)
\be  M_5^2=
\lim_{R_5\rightarrow\infty}R_5\, M^2_4(R_5)
=\lim_{R_5\rightarrow\infty}e^K|Se_s+Te_t+Ue_u|^2R_5
=|se_s+te_t+ue_u|^2\ ,
\en
in agreement with eq.(\ref{ze}).

The large radius limit is different however for states associated to singular
points in the moduli space for which $e^{K}$ does not fall off like $R_5^{-3}$.
This happens for massive hypermultiplets which become massless
at the conifold points \cite{s2}. Their 4-D mass is given by
\be
M_4^2=e^K|Z|^2\ ,\label{m4}
\en
which goes to zero in the $Z\rightarrow 0$ conifold limit.
Near $Z=0$, the prepotential behaves as ${\cal F}\sim iZ^2\ln Z$.
The corresponding K\"ahler potential,  after setting
$Z=zR_5$, behaves as
\be
e^{-K}=[2z\zbar\ln z\zbar+2z\zbar\ln R_5^2-(z-\zbar)^2]R_5^2
\sim 2z\zbar R_5^2\ln R_5^2\ .
\en
Using eq.(\ref{m4}) we obtain
\be
M_4^2\sim \frac{1}{\ln R_5}\ ,\label{m44}\en
so that
\be  M_5^2=
\lim_{R_5\rightarrow\infty}R_5\, M^2_4\sim {R_5\over\ln R_5}\rightarrow\infty\
{}.
\label{m55}\en
We see that although the 4-D mass goes to zero in the large radius
limit, the 5-D mass diverges, therefore these states are not
present in the decompactified theory. This is expected from the
fact that they are due to the world-sheet instanton effects which
arise from the mirror map, but such effects are not present
in five dimensions as mentioned in section 2.

On the other hand, massive vector multiplets which never become
massless in $D=4$, keep a finite mass in the 5-D decompactification limit.
{}Furthermore, two massless vector multiplets appear at $R=1$, enhancing
one of the $U(1)$'s to $SU(2)$. This can be compatible with the type IIA
description if we accept the existence of enhanced symmetry points
on Calabi-Yau threefolds in the large volume limit, for $t^1=t^2$
\ie\ at $R=1$. A similar phenomenon occurs in the case of the heterotic
superstring compactified on $T_3$ which is dual to 11-D
theory compactified on $K_3$, where enhanced symmetry points
do indeed exist \cite{w,branes,asp2}. As shown in refs.\cite{w,asp2}, the
enhanced symmetry points of the Narain lattice correspond to rational curves
which shrink to zero size
\ie\ orbifold points on the $K_3$ side. A necessary condition for
the existence of such points is the vanishing of
the two-index antisymmetric tensor field. This is automatic in 11-D
supergravity
with $p$-branes since there is no two-form wrapping a complex curve.
It follows that the dual pair consisting of 11-D theory compactified on $K_3$
and heterotic superstring compactified on $T_3$ is fully described
by classical physics of the $K_3$ side.
In the case of CY threefolds in $D=11$ we are in a similar situation since
there is no two-index antisymmetric tensor field, hence
there are no ``instanton'' effects \cite{cf,bbs}. Here again,
massless vector multiplets do appear, reflecting the underlying $K_3$ fibration
of the Calabi-Yau manifold. The description  of our dual pair as a classical
Calabi-Yau compactification of 11-D theory should remain exact. Note that in a
more general case, massless charged hypermultiplets could also exist at the
enhanced symmetry points. Their vacuum expectation values would connect
Calabi-Yau threefolds with distinct topologies in analogy with the 4-D example
of ref.\cite{asp1}.

The central charge formulae (\ref{ze}-\ref{z3}) indicate that further
enhanced symmetries (and/or massless states) may be present
for other values of $t^1,t^2$.\footnote{These could be related to
non-perturbative enhanced symmetries recently discussed in ref.\cite{w2}.} In
particular, massless states could appear which are charged with respect to
$b_{\mu\nu}$. They would induce non-perturbative modifications of the Yukawa
couplings, possibly generating $\cal V$-terms that are also quadratic and/or
cubic in $s$. In the next section we will show that such points do indeed exist
in the strong coupling regime of 5-D heterotic theory.

\section{Strong Coupling Regime of 5-D Heterotic Theory}

In order to derive the effective heterotic action in the strong coupling
regime, we will first compactify the 5-D model to $D=4$ on a circle of radius
$R_5$. By using duality we know that the exact theory is described by type II
superstring compactified on an appropriate CY threefold. Then, we go back to
$D=5$ by taking the limit $R_5\to\infty$ in a way that corresponds to the
strong coupling of the heterotic model. We will illustrate this procedure
on the two-moduli ($ST$) example which on the type II side
correspond to $X_{12}(1,1,2,2,6)$ CY model.

This model contains, at least in the weak coupling region,
the antisymmetric tensor multiplet coupled to $N=2$ supergravity
in five dimensions.\footnote{It contains also 129 hypermultiplets which
are irrelevant to the following discussion.}
The effective action (\ref{lagr}) is described in this region exactly by the
function
\be
{\cal V}=st^2 +b\, t^3\ , \label{stt}
\en
where $b$ is a constant.
The $t^3$ term is unphysical at the perturbative level since
it can be removed by shifting $s$ as in eq.(\ref{sshift}).

Eq.(\ref{stt}) can be recovered from the decompactification limit of the
corresponding $D=4$ rank 3 model, with the 4-D moduli fields identified as
\be
T=itR_5\qquad\qquad S=i{R_5\over t^2}\ .\label{ST}
\en
The weak coupling region is defined by $\makebox{Im}S>\makebox{Im}T$
which corresponds to $t<1$.
This implies that when $R_5\to\infty$ the limit $S\to i\infty$ should be taken
first to recover the weakly coupled heterotic model in $D=5$. In fact as
$S\to i\infty$, the heterotic prepotential has the form
\be
{\cal F}=ST^2+f(T)+{\cal O}(e^{2i\pi S}) \label{prep}
\en
where $f$ is the one loop correction of ref.\cite{stt}. $f$ is defined up to a
quartic polynomial with real coefficients which can be removed by a symplectic
change of basis that is also a symmetry of the perturbative theory.
Furthermore,
it is easy to show that $\partial_T^5f\to 0$ as $T\to i\infty$. It follows that
in the decompactification limit $\cal V$ is given by eq.(\ref{stt}) in the weak
coupling region $t<1$.

To find the decompactification limit in the full range of the coupling $t$,
we consider the exact prepotential of the dual type II model
$X_{12}(1,1,2,2,6)$ \cite{kv}. The Yukawa couplings are given by the following
expression, as functions of the $N=2$ special coordinates $T_1$ and $T_2$
\cite{cand,hk}:
\be
{\cal F}_{ijk}= {\cal F}^0_{ijk} + \sum_{0\leq n_1,n_2\in\bf{Z}}
\frac{n_in_jn_k N(n_1,n_2) q_1^{n_1} q_2^{n_2}}{1-q_1^{n_1} q_2^{n_2}}\ ,
\label{yukawa}
\en
where $q_{i}=\exp(2i\pi T_i)$ and ${\cal F}^0_{ijk}$ are the intersection
numbers with ${\cal F}^0_{111}=4$, ${\cal F}^0_{112}=2$ and ${\cal F}^0_{122} =
{\cal F}^0_{222}=0$; $N(n_1,n_2)$
are the instanton numbers, the first few of which have been explicitly given in
ref.\cite{cand}. In order to relate type II to its dual heterotic theory, the
special type II coordinates $T_1$ and $T_2$ must be mapped to the special
heterotic coordinates $S$ and $T$. The perturbative tests of duality
\cite{kv,stt} dictate $T_1=T$ and $T_2=S+\alpha T$, where $\alpha$ is an
arbitrary constant. In ref.\cite{klm} it has been argued that
$\alpha=-1$ based on the physical requirement that the non-perturbative
monodromy transformation $T_1\rightarrow T_1+T_2, ~T_2\to -T_2$ preserves the
positivity of Im$S$ \ie\ of the inverse square of the coupling constant.

Let us consider first the decompactification limit $S,T\to i\infty$ in the weak
coupling region $\makebox{Im}S >\makebox{Im}T$.
In such a limit the instanton sum of eq.(\ref{yukawa}) vanishes and we obtain:
\be
{\cal V}=st^2-{1\over 3}t^3\qquad\qquad (t<1)\ , \label{stt1}
\en
in agreement with eq.(\ref{stt}) with $b=-1/3$. Note that the perturbative
symmetry (\ref{sshift}) is broken by non-perturbative
effects to a quantized dilaton shift $s\to s+nt$ with $n$ integer. Hence
the $t^3$ term of eq.(\ref{stt1}) cannot be removed at the non-perturbative
level.

In the strong coupling region $\makebox{Im}T>\makebox{Im}S\to\infty$,
the instanton sum of
eq.(\ref{yukawa}) becomes
\be
-\!\!\!\!\!\sum_{0\leq n_1<n_2\in\bf{Z}}n_in_jn_k N(n_1,n_2) \label{sum}
\en
{}For $n_1\ge 1$ the instanton numbers $N(n_1,n_2)$ vanish for $n_1<n_2$ while
$N(0,n_2)=2\delta_{1n_2}$ \cite{cand}. Hence, only ${\cal F}_{222}$
receives a non vanishing contribution $-2$ from the instanton sum. This result
can also be obtained from the small $q_1$ expansion of the Yukawa couplings
given in eq.(5.7) of ref.\cite{ly}:
\be
{\cal F}_{222}={2q_2\over 1-q_2} +{\cal O}(q_1)\ .\label{ely}
\en
It follows that
\be
{\cal V}=s^2t-{1\over 3}s^3\qquad\qquad (t>1)\ .\label{stt2}
\en
Note that eq.(\ref{stt2}) is the same as eq.(\ref{stt1}) with $s$ and $t$
interchanged reflecting a non-perturbative symmetry.

The final result for the function $\cal V$ in the two-moduli model,
eqs.(\ref{stt1},\ref{stt2}), can be written as
\be
{\cal V}=[st^2-{1\over 3}t^3]\theta(1-t) +[s^2t-{1\over 3}s^3]\theta(t-1)\ .
\label{stt3}
\en
As mentioned in section 3, this model has no enhanced gauge symmetry point in
the weak coupling regime. The conifold singularity which is present in four
dimensions at $T=i$ disappears upon decompactification to $D=5$. On the other
hand a discontinuity appears in the exact theory at $t=1$ which corresponds in
four dimensions to the non-perturbative singularity at $S=T$ ($q_2=1$, \cf\
eq.(\ref{ely})). This is a fixed point of the non-perturbative monodromy
transformation $S\leftrightarrow T$ which exchanges the heterotic string
coupling constant with the compactification radius. Following the discussion of
section 3, the singularity at $t=1$ must be due to solitonic excitations
which become massless at this point in  five dimensions. Their existence is
clearly a generic feature of the 5-D theory.

In fact, the above analysis can be extended
in a straightforward way to the rank 3 model studied in section 3 which
upon compactification to $D=4$ becomes dual to type II superstring
on $X_{24}(1,1,2,8,12)$ CY threefold. Here again, in the
strong coupling region $(2\phi)^{-3/2}>R>(2\phi)^{3/2}$
[$\makebox{Im}S<\makebox{max}\{\makebox{Im}T,\makebox{Im}U\}$]
one finds $\theta$-function discontinuities which are due
to non-perturbative states that become massless at $S=T$ and $S=U$.

In conclusion, there is a correspondence between the singularity structure of
moduli spaces of 5-D and 4-D theories which comes out very clearly
from our analysis. In $D=4$, the enhanced symmetry points generically disappear
for finite values of the heterotic coupling constant, being replaced by
conifold
singularities. In $D=5$, enhanced symmetries survive non-perturbative effects
and conifold singularities are absent. The additional singularities whose
existence is a generic feature of CY compactifications
have a very clear interpretation in $D=5$. They are due
to massless non-perturbative states which manifest their presence through
discontinuities of the effective action. It would be interesting to determine
what are the quantum numbers of these states.

\noindent{\bf Acknowledgments}

We are grateful to R. Khuri, W. Lerche, P. Mayr and especially
to A. Klemm for very useful conversations. I.A. thanks the Department of
Physics at Northeastern University for its kind hospitality. S.F. and T.R.T.
acknowledge the hospitality of Aspen Center for Physics during initial stage of
this work.

\end{document}